# Impact of cation redox chemistry on continuous hydrothermal synthesis of 2D-Ni(Co/Fe) hydroxides

**Authors**: Massimo Rosa[1*], Debora Marani[2], Giovanni Perin[3], Søren Bredmose Simonsen[1], Philipp Zielke[1], Antonella Glisenti[3], Ragnar Kiebach[1], Andreas Lesch[4] and Vincenzo Esposito[1*]

[1]*DTU Energy, Technical University of Denmark, Risø Campus, Frederiksborgvej 399, 4000, Roskilde, Denmark*

[2]*Centro de Engenharia, Modelagem e Ciências Sociais Aplicadas, Universidade Federal do ABC, Santo André, (SP) 09210-580, Brazil*

[3]*University of Padova, Department of Chemical Sciences, via Marzolo 1, 35131, Padova, Italy*

[4]*Laboratoire d'Electrochimie Physique et Analytique, Ecole Polytechnique Fédérale de Lausanne, EPFL Valais Wallis, Rue de l'Industrie 17, CH-1950 Sion, Switzerland*

*Corresponding authors: Massimo Rosa, masros@dtu.dk, Vincenzo Esposito, vies@dtu.dk.

**Abstract:** Continuous hydrothermal flow synthesis (CHFS) is a *facile*, upscalable and cost-efficient synthetic method enabling the nanostructuring of advanced functional materials in steady conditions, *i.e.* not in batch synthesis. In this paper, we use CHFS to crystallize NiCo- and NiFe-Hydroxides in water solution with 2D nanofeatures. By tuning the synthetic parameters, we disclose the key role of the cation redox chemistry in the transition between two competitive phases: from 2D-nanoplatelets of brucite to layered double hydroxides (LDH). For controlling the precipitation of different Ni,Fe,Co-Hydroxide phases, we propose the combined use of an oxidizing ($H_2O_2$) and a complexing agent ($NH_3$). At temperatures as low as 80 °C, the presence of $H_2O_2$ and a low concentration of $NH_3$ favour the $Ni^{2+}/Co^{3+}$ over $Ni^{2+}/Co^{2+}$ oxidation states, shifting the product structure from brucite phase (temperatures > 80 °C) to LDH. Conversely, for the NiFe-Hydroxides the transition from LDH (temperatures ≤ 80 °C) to brucite phase (temperatures > 80 °C) is controlled by the reaction temperature only. Due to the high stability of $Fe^{3+}$, the synthesis of NiFe products by CHFS does not require oxidizing and complexing agents, resulting in a robust process for large-scale production.

## 1. Introduction

Nanostructured hydroxides of Ni, Co, and Fe are currently attracting great attention for their catalytic activity towards the Oxygen Evolution Reaction (OER) as an alternative to state-of-the-art Ir- and Ru-based oxides [1–5]. While scarcity and high cost of Ir and Ru limit their application at the industrial scale, Ni/Co/Fe- based hydroxides are largely available and exhibit outstanding catalytic performances in terms of both activity and durability [6–12]. However, nanostructuring of functional hydroxides is challenging due to the high-density polymorphism with a variety of properties and competing phases, e.g. for iron hydroxides [13].

NiCo- and NiFe-Hydroxides mainly exist with two different crystalline arrangements, depending on the oxidation state of the metal cations. When only divalent cations, i.e. $Ni^{2+}/A^{2+}$ (A = Co), are incorporated, the hydroxides exhibit a brucite-like phase. This consists of two-dimensional (2D) sheets made of edge sharing octahedra with the divalent metal cation coordinated by six hydroxide groups. The 2D metal-hydroxyl sheets stack via hydrogen bonding to form a closely-packed three-dimensional (3D) network [14].When trivalent cations substitute isomorphically the divalent ones with similar ionic size, i.e. $Ni^{2+}/A^{3+}$ (A = Fe, Co), the 2D layers acquire a positive charge that is compensated by the incorporation of exchangeable anions into the interlayer galleries [7].The latter compounds are referred to as layered double hydroxides (LDHs) and their general formula is $M^{II}_{1-x}M^{III}_{x}\ (OH)_2(A^{n-})_{x/n}$, where $M^{II}$ and $M^{III}$ are the divalent and trivalent cations and $A^{n-}$ the intercalated anion. The structure of LDHs is equivalent to anionic clays of the hydrotalcite type [15] consisting of stacked 2D-nanosheets prepared using different metals both at the $M^{II}$ site ($Ca^{2+}$, $Mg^{2+}$, $Ni^{2+}$, $Co^{2+}$, $Mn^{2+}$, etc.) and $M^{III}$ site ($Al^{3+}$, $Fe^{3+}$, $Co^{3+}$) [16]. Their versatility in terms of composition, anion exchange and the possibility of exfoliation make them attractive for a wide range of industrial and technological applications, such as catalysis, catalyst support, anionic adsorption, stabilizers and flame retardants in functional polymers [17].

Hydroxide-layered materials are typically synthesized using well-known methods, such as solvo/hydro-thermal [8,9,18–20] and co-precipitation approaches [21–23].

For instance, in the case of NiCo LDHs, the hydrothermal route is preferred due to the instability of $Co^{3+}_{(aq)}$ that hinders the direct use of Co(III) precursors. Therefore, the reaction is carried out in presence of ammonia or an amine in order to oxidize the complexed Co(II) to a stable Co(III) amine complex by means of dissolved oxygen [24]. The starting solution is generally heated at 90-160 °C and kept from 3 to 14 hours, during which Co(II) is oxidized and the Co(III) and Ni(II) aminocomplexes decompose to the corresponding hydroxides, forming the LDH phase [18,19,25].

In contrast, the synthesis of NiFe LDHs does not require manipulating the oxidation states of the cations due to the higher stability of $Fe^{3+}_{(aq)}$ compared to $Co^{3+}_{(aq)}$. Typical approaches consist of both hydrothermal synthesis [8,10,26] and coprecipitation processes [21–23,27] where $Ni^{2+}_{(aq)}$ and $Fe^{3+}_{(aq)}$ are slowly added to a $OH^-$ solution containing the desired intercalating anion. Similarly to the hydrothermal synthesis of NiCo LDH, the product is collected after several hours as a result of the slow dropwise addition and/or ageing of the mixed solution [8,9,21,22,26,27].

Despite solvo/hydro-thermal and coprecipitation synthesis being undoubtedly versatile methods, they suffer from drawbacks related to the discontinuity and duration of a batch process. During the synthesis, products and/or reactants continuously evolve with time leading to a limited control of the supersaturation conditions, and eventually particle growth. In the case of solvo/hydrothermal synthesis, the concentration of the reactants varies according to the reaction kinetics, so that particles nucleated at different times experience different compositions of the precursor solutions into the reactor. In the case of coprecipitation, particles formed in equilibrium conditions age and grow for longer times, broadening the particle size distribution [28]. These difficulties in providing a constant reaction environment make the scalability of batch processes challenging.

As consequence, applying these methods to an industrial scale, with high economic and environmental costs, could make them prohibitive.

An effective approach to overcome such limitations is the continuous hydrothermal flow synthesis (CHFS) method, where streams of reactants are continuously mixed through a reactor at desired temperature and pressure [29]. In CHFS, the precursors experience steady-state physicochemical conditions within the reactor for a controlled reaction time. Indeed, the residence time is finely tuned to obtain constant and highly reproducible materials in terms of morphological, structural, and compositional features. The process is typically fast, with residence times lying between a few seconds to a minute. This leads to high throughput nanomaterials production with a superior control on the synthesis parameters [30-32], enabling for an efficient upscalable process readily applicable at industrial level of production. In this vein, CHFS has been used for the synthesis of a wide range of metal oxide and hydroxide nanoparticles [33], even at the industrial level [34]. Indeed, the method offers undoubted advantages in terms of highly reproducibility of products with well-defined properties and the readily scalability of the whole process.

CHFS has been also investigated for the synthesis of LDH with various compositions [28,35,36]. For the NiCo LDHs, Liang et al. have recently investigated particles and morphology formation by means of a small continuous flow reactor [37]. However, this system employed flow rates as low as 1 mL $min^{-1}$, resulting in a residence time of ca. 30 minutes. This makes the system not representative of a large-scale CHFS reactor. On the other hand, Darr's group investigated the production of NiCo-Hydroxides in a CHFS reactor with a much higher flow rate of 20 ml $min^{-1}$, but only focused on the preparation of brucite phases with different Ni : Co ratios [38]. However, an analysis of NiFe- and NiCo-LDH synthesis via CHFS remains mostly unexplored.

In this work, we studied the synthesis of NiCo- and NiFe-Hydroxides focusing on the CHFS process conditions for analysing the transition between brucite-like and LDH phases. Due to high industrial relevance, we especially focus on fast synthesis, i.e. the residence time limit < 1 min with flow rates as high as 40 mL $min^{-1}$ [37], yielding out-of-equilibrium conditions, where the material produced strongly depends on the reaction kinetics. In particular, we investigate the role of the different redox chemistries of Fe and Co, pointing out remarkable differences compared to batch processes emerging from the operation under kinetic control.

## 2. Experimental

### 2.1 Materials

For the hydrothermal syntheses, analytical grade $Co(NO_3)_2 \cdot 6\ H_2O$ , $Ni(NO_3)_2 \cdot 6\ H_2O$, $Fe(NO_3)_3 \cdot 9\ H_2O$, KOH and $K_2CO_3$, (all purchased from Sigma-Aldrich) were used as received. For the synthesis of the NiFe-based compounds, KOH was added to the reaction solution to increase the pH. A $NH_3$ solution (28 wt.%, VWR chemicals) was employed as complexing agent and $H_2O_2$ 30% wt (Sigma-Aldrich) as oxidant in the syntheses of NiCo products.

### 2.2 Hydrothermal reactor

A detailed description of the CHFS setup has been given in a previous publication [39]. Briefly, **Figure 1** represents a 3D rendering of the reactor highlighting the most important components. The system consists of five piston pumps controlling the flows of liquid streams introduced through five different inlets (F1-F5). F1 represents the heating medium of the reactor and consists of a stream of deionized (DI) water, which is pumped through the first heater. The stream of hot water is subsequently mixed with reactant streams F2 and F3 at the mixing stage M1 and with F4 and F5 at the second mixing stage M2. The final mixed stream flew through a second heater (re-heater in Fig. 1) that prevented excessive cooling of the resulting solution after mixing with the cold streams F2-F5. The produced particle dispersion is eventually cooled down through a heat exchanger and is depressurized before collection.

**Table 1** provides all experimental conditions that are herein applied for each synthesis of the NiCo materials, while **Table 2** reports the conditions for the NiFe-hydroxide products. In all NiCo-Hydroxide experiments, a cation ratio Ni:Co 2:1 was used, while for NiFe-hydroxide compounds the ratio was 3:1. These stoichiometries were selected to investigate the influence of the CHFS process on compositions that can form stable LDH phases [7]. In addition, for NiCo hydroxides, 2:1 corresponds to the cation ratio used by Liang et al. [37] in the only attempt of synthesizing NiCo LDH with a continuous process. In the case of NiFe LDH, no previous work on the continuous synthesis is known. However, there are evidence indicating a higher stability of the 3:1 composition compared to stoichiometries with a higher content of Fe [40]. The products from the different experiments are labelled NiCoX or NiFeX where X corresponds to the sequential number of the experiment, see Table 1. Each table specifies the reactor pressure and the temperatures measured during the syntheses by using three thermocouples within the reactor, T1, T2 and T3 (see Figure 1). These sensors monitored the temperature of F1 after the first heater (T1), the temperature of the mixed stream after M2 (T2), and the temperature after the second heater (T3). The two tables also report the flow of the different streams pumped into the reactor and their compositions. **Table 2** shows that for the synthesis of NiFe-hydroxides, the metal cations and KOH were added through different inlets (F4 and F2 respectively) to prevent precipitation of the hydroxides before the mixing stage M2. Differently, for NiCo-Hydroxides (Table 1), $NH_3$ was added both simultaneously or separately as ammonia complexes to prevent the cations precipitation in the precursor solution. The residence time of the reactants at the hydrothermal conditions (last column in Table 1 and 2) is calculated from the applied flows and the sizes of the reactor pipes.

2.3 Characterization

After synthesis, the particles were washed by centrifuging the material and redispersing it in fresh DI water several times. The products were dried at 70 °C to obtain powders for the diffractometric analysis. X-ray diffraction measurements were carried out with a Bruker D8 (Cu Kα radiation) in a *θ/2θ* configuration with 0.025° step size and 2 s step time. The reference diffraction pattern for $Ni(OH)_2$ (JCPDS-14-0117) and $Co(OH)_2$ (JCPDS-30-0443) were taken from the standard JCPDS cards. “The lattice parameters were estimated considering the hexagonal crystal structure for brucite products, with the lattice parameters following the conditions a=b≠c (and α=β=90° and γ=120°). Therefore, the standard formula for the hexagonal system was applied using the crystal planes of (001) and (100) of the brucite phase [41]:

$$\frac{1}{d_{hkl}^2} = \frac{4}{3}\left(\frac{h^2 + hk + k^2}{a^2}\right) + \frac{l^2}{c^2} \qquad \text{Equation 2.1}$$

where $d$ is the interplanar spacing, $h$, $k$, $l$ represents the Miller indices of the planes and $a$, $b$ and $c$, the lattice parameters. The spacing corresponding to the planes (101) and (200) of the anatase phase were determined by applying the Braggs´s equation:

$$d_{hkl} = \frac{n\lambda}{2\, sen\theta} \qquad \text{Equation 2.2}$$

where $d$ is the spacing of the crystal planes, $(h, k, l)$ are the Miller indices, $n$ is a positive integer, λ is the wavelength of the incident X-ray radiation (0.154 nm), θ is the Bragg´s diffraction angle.

The unit cell volume was calculated applying the following equation [40]:

$$V_{cell} = a^2 c \sin 60^\circ \qquad \text{Equation 2.3}$$

X-ray photoelectron spectroscopy (XPS) was performed by means of a Perkin Elmer PHI 5600 ci Multi-Technique System, using a Mg Kα anode (energy: 1253.7 eV) working at 250 W. The spectrometer was calibrated by assuming the binding energy (BE) of the Au $4f_{7/2}$ line to be 84.0 eV with respect to the Fermi level. Extended spectra (survey - 187.85 eV pass energy, 0.5, 0.05 s/step) and detailed spectra (for Ni 2p, Co 2p, Fe 2p, O 1s and C 1s - 23.5 eV pass energy, 0.1, 0.1 s/step) were collected, with a corresponding standard deviation of the BE values of 0.10 eV. The atomic percentage, after a Shirley-type background subtraction was evaluated by using the PHI sensitivity factors [42,43]. The peak positions were corrected for the charging effects by considering the C 1s peak at 285.0 eV and evaluating the BE differences.

Differently, for the microscopy analyses, the purified dispersions were drop casted on TEM grids and metallic stubs. HRTEM imaging, SAED measurements and STEM/EDS elemental analyses were performed using two different TEM microscopes: a JEOL 3000F microscope equipped with an Oxford Instrument detector for the EDS analysis operated at 300 kV, and a Tecnai Spirit operated at 120 kV. The elemental compositions reported in Table 4 were calculated from EDS data collected with a Zeiss Merlin SEM equipped with a Bruker XFlash 6 EDS detector.

## 3. Results and Discussion

Different process variables (*i.e.* temperature, complexing agent and oxidizing agent amounts) are investigated to explore the capabilities of the CHFS method in enabling the shift from brucite to hydrotalcite products. Specifically, we address the versatility of the developed synthetic approach in two different synthesis: (i) NiFe- and (ii) NiCo-Hydroxides. While the former involves cations with stable oxidation states in water ($Fe^{+3}$), the latter is affected by the equilibria between Co(III) and Co(II) cations and their complexes.

However, regardless of the specific reaction conditions, the synthesis of nickel, iron and cobalt-based layered materials occurs via an out-of-equilibrium condition CHFS method, achieved by keeping the flow rate as high as 40 mL $min^{-1}$. These conditions result in a residence time below 1 minute that, while providing a constant reaction environment, can maximize the effect of the reaction kinetics on morphology and crystallographic structure of the final layered product.

**Figures 2** show the XRD patterns for Ni,Fe,Co-Hydroxides and crystal parameters for brucite phase. Specifically, Figure 2(A) and Figure 2(B) show NiFe-Hydroxides and NiCo-Hydroxides obtained at different synthetic conditions, respectively; Figure 2(C) and Figure 2(D) summarize the cell volume and the crystal parameters (*a* and *c*) calculated from XRD patterns (reported in Table 3).

For NiFe-Hydroxides (Figure 2(A)), two samples, NiFe1 and NiFe2, are prepared at high and low reaction temperatures, respectively. Specifically, NiFe1 crystallizes at T2/3 150 °C with water heated at T1 296 °C, whereas NiFe2 forms at T2/3 80 °C with water heated at T1 156 °C.

For NiFe-Hydroxides, the synthesis is conducted without adding either a complexing agent (*i.e.* $NH_3$) or an oxidant (*i.e.* $H_2O_2$), and iron is directly added as $Fe^{3+}$ due to its stability in water solution. As indicated in the XRD patterns in Figure 2(A), the product NiFe1 shows a brucite-like phase. In particular, the peak positions, the unit cell volume and parameters of product NiFe1 are similar to those corresponding to the $Ni(OH)_2$ crystallographic cell (see Figure 2(C) and 2(D)). These results likely suggest that NiFe1 consists of brucite-like $Ni(OH)_2$ and of an amorphous Fe-based compound that does not generate any XRD peaks (likely amorphous FeO(OH)).

On the other hand, NiFe2 clearly displays a hydrotalcite-like structure, typical of LDH materials. Such a result indicates a clear effect of the temperature in favouring either the formation of the brucite-like product at the highest temperatures (T2/3 150 °C) or the hydrotalcite-like (LDH) structure at the lowest temperatures (T2/3 80 °C). The observed broadening and weak intensity of the peaks for both products can be ascribed to a different number of contributions such as nanometric lateral size and shape anisotropy of crystals, intercalation of different species ($NO_3^-$, $CO_3^{2-}$, $H_2O$) [44,45] and presence of different LDH polytypes [15]. Interestingly, the peaks with Miller index l = 0 are sharper (*e.g.* 100 and 110) indicating a higher crystalline order in-plane, compared to through-plane directions where $l \neq 0$. This observation also suggests the presence of a stacking disorder compatible with the presence of intercalating species with different size, which induces variations in the plane spacing.

As for NiCo-Hydroxides, the instability of $Co^{3+}$ in water prevents its direct use in the reactor. Therefore, the experiments employ $Co^{2+}$ and subsequently promote its *in situ* oxidation. In this synthetic route, we add an excess of ammonia as ligand to complex $Co^{2+}$ and $Ni^{2+}$. This avoids the precipitation of the hydroxides in the precursor solutions and supports the oxidation of $Co^{2+}$ [46] according to the reaction below:

$$4\ Co(NH_3)_6^{2+} + O_2 + 2H_2O \rightarrow 4\ Co(NH_3)_6^{3+} + 4OH^-$$

The presence of ammonia also provides an alkaline pH required for the hydroxide synthesis. The different stability of amino complexes at different reaction conditions makes the identification of the transition from brucite to hydrotalcite for NiCo-Hydroxides a complex case, that has required a larger number of experiments (from NiCo1 to NiCo9) compared with Fe system (only 2). Therefore, a number of NiCo-Hydroxides (9 samples) are produced to systematically explore the conditions enabling the transition from brucite-like to LDHs phase. Figure 2(B) reports the XRD patterns for the synthesized products, indicating the formation of brucite-like phases consisting of a solid solution of $Ni(OH)_2$ and $Co(OH)_2$ at temperatures varying in the 85-161 °C range (T2; Table 1) and in the presence of an excess of $NH_3$ (NiCo1-NiCo5). Interestingly, lowering the temperature of the mixed stream (T2) from the highest investigated value (161 °C in NiCo2) to around 85 °C (in NiCo4) produces a shift of the peaks towards lower angles, characteristic of the brucite-like structure of pure $Co(OH)_2$. The variation of the peaks' positions suggests the formation of solid solutions with different compositions, despite an identical cations' ratio in the precursors solution (Ni:Co=2:1). Specifically, Ni-rich compositions can be inferred at high temperature for the sample NiCo1, NiCo2 and NiCo3, whereas the two low temperature products, NiCo4 and NiCo5, are likely Co-rich product (see Figure 2(C) and 2(D)). This effect can be ascribed to a different availability of the cations due to a greater stability of the $Ni^{2+}$ amino complexes compared to $Co^{2+}$ amino complexes towards hydrolysis at NiCo4 and NiCo5 reaction conditions. The formation of $Co(NH_3)_6^{2+}$ at standard conditions indeed has logK = 5.11 [42], while for $Ni(NH_3)_6^{2+}$ logK = 8.3 [47] indicating that the $Ni^{2+}$ amino complex is three order of magnitudes more stable than the $Co^{2+}$ one. Such a hypothesis further relies on the observation that the supernatant collected from the reactor contains some residual precursor, likely $Ni^{2+}$ amino complex. Being $Ni(NH_3)_6^{2+}$ more stable than $Co(NH_3)_6^{2+}$, a larger availability of $Co^{2+}$ is expected at low temperature, leading to materials with an XRD pattern compatible with Co-rich structures.

As in the case of NiFe-Hydroxides, the observed slight differences in the broadening of the crystallographic peaks are not necessarily associated with different level of crystallinity, but rather with a number of contributions such as shape anisotropy of crystals.

Interestingly, with respect to the $Ni(OH)_2$ (JCPDS 14-0117, $a_0$= $b_0$=3.126 Å and $c_0$=4.605 Å), the introduction of $Co^{2+}$ leads to a larger distortion of the unit cell along the two equal axes (a and b) rather than along the c direction (height of the unit cell). Indeed, while a progressive and moderate increase of the c parameter values along NiCo1-NiCo5 series is observed, an abrupt variation in the a parameter values (and in the unit cell volume) between NiCo1-NiCo3 and NiCo4-NiCo5 products is instead obtained. This empirical result is clearly associated with the transition from Ni-rich to Co-rich brucite materials, characterized by unit cell parameter values closer to those describing $Co(OH)_2$ (JCPDS 30-0443, $a_0$= $b_0$=3.183 Å and $c_0$=4.652 Å). Yet, within each series, NiCo1-NiCo3 and NiCo4-NiCo5, the variation in the value of the cell parameter a, is minimal and within the experimental error (< 1%).

In NiCo5, the temperature is further lowered by introducing the precursor solution in the second mixing stage (M2), therefore avoiding direct mixing with the hot water stream. These conditions promote the formation of a detectable fraction of hydrotalcite-like phase together with the brucite phase, as indicated in the XRD pattern of NiCo5. We can explain the presence of this secondary hydrotalcite phase as a stabilization effect of $Co^{3+}$, formed in the precursor after mixing with $NH_3$, by the temperature reduction. The effect of the temperature on the oxidation state of cobalt and stability of the corresponding amino complexes is then verified by adding $NH_3$ (complex agent) to a water solution of $Co^{2+}$ at 25 °C. Upon the addition of $NH_3$, the solution turns from pink to brown suggesting the presence of $Co^{3+}$ species [48,49] in the form of e.g. $Co(NH_3)_6{}^{3+}$ [34] or $[Co(NH_3)_5O_2Co(NH_3)_5]^{4+}$ [24,50]. When the temperature is increased to 90 °C, the solution turns red and the precipitation of a black material corresponding to $Co_3O_4$ (XRD not shown here) is observed. This result indicates the simultaneous presence of $Co^{2+}$ and $Co^{3+}$ at 90 °C, and the tendency of the system to favour the $Co^{2+}$ species at the highest temperatures.

To support the oxidation of $Co^{2+}$ to $Co^{3+}$, hydrogen peroxide ($H_2O_2$) is used in the experiment NiCo6. The addition of $H_2O_2$ to a high temperature process (142 °C at T2) leads to a mixture of different phases, as indicated by the XRD pattern. However, none of the observed peaks correspond to a hydrotalcite-like phase. As the simple addition of $H_2O_2$ in NiCo6 does not form an LDH product, NiCo7 combines the presence of the oxidizing agent with a low temperature process (83 °C at T2) in order to enhance the formation of $Co(NH_3)_6{}^{3+}$ in the precursor and avoid its reduction at T > 85 °C. Unexpectedly, we do not observe any precipitation of a solid product for such an experiment. This is ascribed to a greater stability of the $Co^{3+}$ amino complexes toward the hydrolysis with respect to the $Co^{2+}$ amino complexes [47]:

$Co^{3+} + 6NH_3 \rightarrow Co(NH_3)_6{}^{3+}$, logK = 35.2

$Co^{2+} + 6NH_3 \rightarrow Co(NH_3)_6{}^{2+}$, logK = 5.11

To promote the hydrolysis of $Co(NH_3)_6{}^{3+}$, while keeping a low temperature process, the experiment NiCo8 increases the residence time of the process from 41 s to 60 s. Despite this modification, the reaction still yields to no solid product.

To reduce the stability of the amino complexes without increasing T, NiCo9 employs a lower concentration of ammonia in the reaction, 4.1 M. This value represents the lowest amount of $NH_3$ for avoiding precipitation of the $Co^{2+}$ hydroxides in the precursors. The process also uses low temperature (82 °C at M2) and $H_2O_2$ as understood from the analysis of temperature and oxidizing agent effects in the previous reactions. This combination results in the disappearing of brucite peaks for NiCo9, with the formation of very broad reflections suggesting an LDH-like phase with a high crystallographic disorder.

Surprisingly, our observations on the cobalt cations redox chemistry are not in line with the data reported in literature. Indeed, whilst the oxidation of $Co^{2+}$ in ammonia at room temperature is well known and hereby confirmed, increasing temperature generally enhances the oxidation of $Co^{2+}$ to $Co^{3+}$[24,49–51]. Conversely, in our study, in the presence of an excess of $NH_3$, $Co^{3+}$ reduces back to $Co^{2+}$ when the T is increased to > 85 °C up to 160 °C. This result has never been reported before.

In summary, the XRD analysis suggests that in our CHFS experiments, $Co(NH_3)_6{}^{3+}$ forms in the precursors, but readily reduces to $Co^{2+}$ in reactions where the mixed stream reaches T > 85 °C. These conditions lead to products with a brucite-like phase consisting of a solid solution of $Ni(OH)_2$ and $Co(OH)_2$. Decreasing temperature and ammonia concentration prevent the reduction of $Co^{3+}$ and promote the hydrolysis of the amino complex leading to a disordered LDH-like phase.

To further investigate the reaction mechanism, EDS analysis is performed on the synthesized products. The data are listed in **Table 3** and match well with the discussion of the XRD data. For the samples having a brucite phase, the chemical formulas, calculated from EDS data and indicated in Figure 3(C) and 3(D), are in agreement with the cell parameters obtained by the analysis of the XRD patterns. Indeed, at high temperatures (161-144 °C) the formation of Co-doped $Ni(OH)_2$ is promoted, having a stoichiometry corresponding to the cation ratio used in the precursor. This result indicates that at high temperature the hydrolysis rates for Ni and Co amino complexes are comparable, leading to a similar availability of the two cations and eventually a stoichiometric product. Differently, at lower temperatures (around 80 °C) the synthesis of Co-rich brucite materials is obtained. As already mentioned, this can be associated with the larger stability of the $Ni^{2+}$ amino complexes towards hydrolysis with respect to $Co^{2+}$ amino complexes at low temperatures [46,47]. Therefore, for T < 85 °C the hydrolysis rate of $Ni(NH_3)_6^{2+}$ is suppressed with a consequent larger availability of $Co^{2+}$ in the reaction environment. NiCo9 shows instead an excess of Ni compared to the precursor, which results from the larger stability of $Co(NH_3)_6^{3+}$ compared to all the amino complexes involved in these reactions [47].

Based on the XRD results, NiFe1, NiFe2, NiCo1, and NiCo9 are selected as representatives of brucite- and LDH-like structures with the two compositions, for investigating the cations' oxidation states by X-ray Photoelectron Spectroscopy. **Figure 3** shows the spectra for Ni 2p (3(A), (C), (E), (G)), Co 2p (3(B), (D)), and Fe 2p (3(F), (H)). In Figure 3(A), (C), (E) and (G), Ni 2p spectra present four peaks: besides the $2p_{3/2}$ (855.7 eV) and $2p_{1/2}$ (873.4 eV) components which are consistent with the presence of $Ni(OH)_2$ [52], the shake-up contributions arising from final state configuration interaction are detected (861.6, 879.4 eV). These signals are fingerprints of the presence of $Ni^{2+}$ oxidation state. Only in the NiFe2 sample, two smaller signals arising from inter-band transitions (867.2, 883.3 eV) are detected [52].

The Co 2p spectra for NiCo1 and NiCo9 samples (Figures 3(B) and (D)) display different features from each other. Indeed, for NiCo1 the analysis reveals that the $2p_{3/2}$ and $2p_{1/2}$ components are centered at 780.8 and 796.6 eV, respectively, in agreement with previously observed values for $Co(OH)_2$ [53]. The presence of $Co^{2+}$ is further supported by the shake-up contributions, which can be detected only in 2+ oxidation state. XPS measurements thus confirm the XRD indications of a brucite product with only divalent cations for NiCo1. Differently, for NiCo9 sample, both peaks of $Co^{3+}$ (centered at 779.7 and 794.8 eV) and $Co^{2+}$ (centered at 780.8 and 796.6 eV) species are observed. The ratio of the areas corresponding to the two oxidation states indicates a surface composition of 35/65 at.% for $Co^{2+/3+}$. Therefore, for the sample NiCo9, XPS analysis points out (and confirms) the synthesis of the LDH structures and it also highlights a possible formation of brucite impurities, which are not detected by XRD.

Differently from NiCo1 and NiCo9, the two NiFe compounds show the same oxidation states for both NiFe1 and NiFe2. In particular, the spectra of Fe 2p show the $2p_{3/2}$ component at 711.6 eV and $2p_{1/2}$ component at 725.3 eV, which are consistent with the presence of $Fe^{3+}$ in the form of both FeOOH and LDHs [54]. The absence of shake-up contributions suggests that no $Fe^{2+}$ is present. Broadened signals have previously been observed for iron in the case of layered hydroxides [54,55]. Therefore, the hypotheses on the formation of brucite $Ni(OH)_2$ together with amorphous FeOOH for NiFe1, and LDHs in NiFe2 are further substantiated.

**Figure 4** reports the spectra of oxygen (O 1s region) for all the four samples investigated. All of them display a wide asymmetric peak arising from the sum of two contributions corresponding to two distinct chemical states [56]. The main contribution is centered at higher energy (531 eV) and is assigned to the hydroxide species, as pure $Ni(OH)_2$ is centered at 531.2 eV [55]. On the other hand, the weaker component (centered at 529 eV) is attributed to the bulk or lattice oxygen and corresponds to oxidic species [56]. The ratio $O^{2-}/(O^{2-}+OH^{-})$ estimates a variation of the $O^{2-}$ amount associated with different chemical environment. The corresponding values are listed in **Table 4**. Interestingly, the samples with the lowest value of ratio $O^{2-}/(O^{2-}+OH^{-})$ (6.1% for NiCo9 and 7.9% for NiFe2) are the two LDHs structures, whereas the highest value (10.3%) is associated with the material obtained from the NiFe1 route and consisting of brucite $Ni(OH)_2$ and amorphous FeOOH.

In summary, the XPS analysis highlights the presence of 65% of $Co^{3+}$ on the surface of NiCo9 particles and 100% of $Fe^{3+}$ on both NiFe1 and NiFe2. These results confirm the formation of LDH-like structures in NiCo9 and NiFe2, and precipitation of a $Fe^{3+}$-containing product in NiFe1, which cannot be incorporated in a brucite phase.

**Figure 5**(A) and (C) show respectively the low and high magnification TEM micrographs for the product NiCo1. In accordance with the sharp peaks identified in the XRD pattern, well-crystallized hexagonal brucite-like 2D platelets can be recognized. Yet, STEM analysis (Figures 5(B) and (D)) confirms the formation of a solid solution of $Ni(OH)_2$ doped with cobalt, which is incorporated as $Co^{2+}$ in the brucite crystallographic phase, as anticipated by the XRD analysis.

On the opposite, the TEM micrograph for NiCo9 (Figure 5(E)) shows extremely small particles with no recognizable crystalline habit. This is expected from the diffraction pattern as broad and low intensity signals are detected. The selected-area electron diffraction (SAED) in Figure 5(F) shows two rings associated with the most intense - though very broad - reflections (012) and (110) of the hydrotalcite structure observed in the XRD patterns. Likely, the limited particle growth and high level of disorder can be the result of low reaction temperature and short residence time preventing the rearrangement of the nucleated structure into a highly crystalline phase.

**Figure 6** shows the morphological analysis via TEM for the sample NiFe2. In Figure 6(A) small and thin platelets of hexagonal shape can be recognized. The selected-area electron diffraction (SAED) inset in Figure 6(A) reports two rings corresponding to the (012) and (110) peaks of an LDH structure, in line with the XRD pattern for the NiCo9 product. The high magnification TEM (Figure 6(B)) illustrates a single platelet with a platelet lateral size of around 50 nm. The SAED inset in Figure 6(B) indicates that the as-produced NiFe2 nano-platelets correspond to single crystal particles. Interestingly, the NiFe2 platelets are smaller with respect to the brucite particles of NiCo1. STEM analysis (Figure 6(C) and 6(D)) confirms that both Ni and Fe are present in the structure. This material, despite the same processing temperature and residence time of NiCo9, exhibits a more developed crystallinity highlighted by the sharper peaks in theXRD patterns and different crystalline habit. This could be ascribed again to the different stability of the cations involved in water. Indeed, the precipitation of NiCo LDHs requires hydrolysis of the $Co(NH_3)_6^{3+}$ introducing an extra step in a very fast and low temperature synthesis. On the other hand, iron is directly added as $Fe^{3+}$, no ligands bind to the cations, which can directly combine into the crystalline structure. The simpler mechanism is also reflected by the elemental data in Table 4. Both NiFe1 and NiFe2 materials show a comparable composition to the starting solutions, in contrast with the range of compositions obtained in the NiCo experiments.

Finally, combining all the data obtained from different characterization methods, the crystalline structure of NiCo1, NiCo9 and NiFe2 can be modelled as in **Figure 7**. In particular, for NiCo-Hydroxides, operating at high temperature in the presence of a large excess of ammonia leads to the formation of a brucite phase without intercalation (Figure 7 (A)). Reducing temperature stabilizes the 3+ oxidation state of cobalt, and lowering the ammonia excess destabilizes the amino complex inducing the precipitation of an LDH structure, Figure 7(B). The same structure is favoured at low temperature in the case of NiFe-Hydroxides, Figure 7(C). However, the synthesis of NiFe LDH does not require the presence of ligands to control the oxidation state of Fe. Figure 7(D) concisely summarizes the results obtained.

## 4. Conclusion

In this work, we explore the effect of temperature and of the oxidizing and complexing agents ($H_2O_2$ and $NH_3$, respectively) in controlling the synthesis of NiCo and NiFe-Hydroxides.

We show that the structure of NiFe-Hydroxides is only determined by the temperature. Specifically, it is demonstrated that the formation of two different phases is favoured at the highest temperatures, i.e. brucite $Ni(OH)_2$ and amorphous Fe(O)OH. On the opposite, we obtain a single NiFe-Hydrotalcite phase at temperature as low as 80 °C.

In contrast, for the NiCo system, we highlight that brucite materials are obtained in a range of temperatures that span from high (161 °C) to low (85 °C), with no addition of $H_2O_2$ and in a large excess of $NH_3$. Lowering the temperature is proven to affect the composition, as confirmed by the shifting from Co-doped $Ni(OH)_2$ at the highest temperatures to Ni-doped $Co(OH)_2$ at the lowest temperatures. In particular, the different stability of the amino complexes ($Ni(NH_3)_6^{2+}$, $Co(NH_3)_6^{2+}$, and $Co(NH_3)_6^{3+}$) determines this trend. At temperatures as low as 80 °C, in the presence of an excess of $NH_3$ and with and without $H_2O_2$, no precipitation in obtained because of the extremely high stability of the nickel(II)/cobalt(III) amino complexes. However, when the amount of $NH_3$ decreases, in the presence of $H_2O_2$, the hydrotalcite NiCo products precipitate.

The different reactivity of the two systems (NiFe and NiCo) is associated with the different stability of $Fe^{3+}$ and $Co^{3+}$ in water solution. These results demonstrate that brucite or hydrotalcite phases with NiCo and NiFe compositions can be produced by CHFS at high flow conditions (40 mL min$^{-1}$) with residence time lower than 1 min, proving the capability of this method for the synthesis of NiFe and NiCo-Hydroxides at a large scale.

## 5. Acknowledgements

This project was partially funded from the Fuel Cells and Hydrogen 2 Joint Undertaking under Grant agreement No 700266. This Joint Undertaking receives support from the European Union's Horizon 2020 research and innovation program and Hydrogen Europe and N.ERGHY. The authors gratefully acknowledge Dr. Paolo Tanchrida for the revision of the English.

## 6. Conflicts of interest

There are no conflicts to declare.

**Tables Captions**

**Table 1**: Conditions for the synthesis of Ni2Co hydroxides.

**Table 2**: Conditions for the synthesis of $Ni_3Fe$ hydroxides.

**Table 3**: Results of the EDS elemental analysis of the produced materials, and cell parameters for the brucite calculated from the XRD patterns.

**Table 4**: $O^{2-}/(O^{2-}+OH^{-})$ ratios calculated from the two components of the O 1s peak measured by XPS for NiCo1, NiCo9, NiFe1, and NiFe2. The hydroxide and oxidic components are centered respectively at 531 and 529 eV.

**Table 1**

| Synthesis of NiCo phases | | | | | | | | | | | | | | | | |
|---|---|---|---|---|---|---|---|---|---|---|---|---|---|---|---|---|
| Sample | T1 [C] $H_2O$ | T2 [C] after M2 | T3 [C] re-heated | p [bar] | F1 [mL/min] | F2 [mL/min] | F3 [mL/min] | F4 [mL/min] | F5 [mL/min] | [$Ni^{2+}$] mmol/L | [$Co^{2+}$] mmol/L | [$NH_3$] M | [$H_2O_2$] mmol/L | [$KNO_3$] mmol/L | [ $K_2CO_3$] mmol/L | Residence Time [s] |
| NiCo1 | 314 | 144 | 160 | 205 | 20.7 | 11.1 | 10.8 | 6.7 | - | 68; F2 | 34; F2 | 4; F3 | - | - | - | 47 |
| NiCo2 | 325 | 144 | 160 | 206 | 20 | 20 | 6.7 | 5.6 | - | 34; F2 | 17; F2 | 5.5; F2 | - | - | - | 45 |
| NiCo3 | 300 | 161 | 162 | 208 | 20 | - | 10.5 | 10.5 | - | 34; F4 | 17; F4 | 5.5; F4 | - | 17; F3 | - | 38 |
| NiCo4 | 153 | 85 | 83 | 201 | 20.4 | 10.4 | - | 9.4 | - | 34; F2 | 17; F2 | 5.5; F2 | - | 20; F4 | - | 61 |
| NiCo5 | 155 | 82 | 83 | 201 | 20.4 | 10.4 | - | 9.4 | - | 34; F4 | 17; F4 | 5.5; F4 | - | - | 20; F2 | 41 |
| NiCo6 | 317 | 142 | 160 | 258 | 20.5 | 10.4 | 10.7 | 7.2 | - | 68; F3 | 34; F2 | 8.9; F2 | 85; F2 | - | - | 48 |
| NiCo7 | 163 | 83 | 84 | 202 | 20.1 | 10.3 | - | 10.2 | - | 34; F4 | 17; F4 | 5.5; F4 | 40; F4 | - | 20; F2 | 41 |
| NiCo8 | 171 | 81 | 82 | 200 | 19.4 | 9.9 | 9.5 | 2.6 | - | 34; F3 | 17; F3 | 5.5; F3 | 40; F3 | - | 20; F2 | 60 |
| NiCo9 | 188 | 82 | 82 | 200 | 19.8 | - | 9.9 | 9.9 | 10.2 | 68; F3 | 34; F5 | 3.1; F5 | 85; F2 | - | 500; F4 | 33 |

**Table 2**

| Synthesis of NiFe phases | | | | | | | | | | | | | |
|---|---|---|---|---|---|---|---|---|---|---|---|---|---|
| Sample | T1 [C] $H_2O$ | T2 [C] after M2 | T3 [C] re-heated | p [bar] | F1 [mL/min] | F2 [mL/min] | F3 [mL/min] | F4 [mL/min] | $[Ni^{2+}]$ mmol/L | $[Fe^{3+}]$ mmol/L | [KOH] mol/L | $[K_2CO_3]$ mmol/L | Residence Time [s] |
| NiFe1 | 296 | 150 | 150 | 197 | 20.3 | 10.7 | - | 10.5 | 75; F4 | 25; F4 | 1; F2 | 20; F2 | 40 |
| NiFe2 | 156 | 80 | 80 | 200 | 20.8 | 10.9 | - | 10.5 | 75; F4 | 25; F4 | 1; F2 | 20; F2 | 37 |

**Table 3**

| Material | Precursor at% ratio Ni:Co(Fe) | Product at% ratio Ni:Co(Fe) | Phase | Lattice Parameters | | |
|---|---|---|---|---|---|---|
| | | | | Unit Cell Volume (Å$^3$) | *a*=*b* (Å) | c (Å) |
| NiCo1 | 66:33 | 66:33 | Brucite | 39.349 | 3.137 | 4.619 |
| NiCo2 | 66:33 | 64:36 | Brucite | 39.233 | 3.132 | 4.619 |
| NiCo3 | 66:33 | 66:34 | Brucite | 39.142 | 3.127 | 4.622 |
| NiCo4 | 66:33 | 23:76 | Brucite | 40.486 | 3.176 | 4.636 |
| NiCo5 | 66:33 | 40:60 | Brucite + LDHs | 40.325 | 3.168 | 4.641 |
| NiCo6 | 66:33 | 57.5:42.5 | Mixed phases | | | |
| NiCo7 | 66:33 | No product | - | | | |
| NiCo8 | 66:33 | No product | - | | | |
| NiCo9 | 66:33 | 70:30 | LDHs | | | |
| NiFe1 | 75:25 | 79:21 | Brucite + amorphous FeOOH | 38.927 | 3.122 | 4.613 |
| NiFe2 | 75:25 | 76:24 | LDHs | | | |

**Table 4**

| Material | $O^{2-}/(O^{2-}+OH^{-})$ ratio |
|---|---|
| NiCo1 | 8.6% |
| NiCo9 | 6.1% |
| NiFe1 | 10.3% |
| NiFe2 | 7.9% |

## Figure Captions

**Figure 1:** Schematic representation of the continuous flow reactor used for the synthesis. F1 to F5 indicate the inlet of the 5 different streams into the reactor (F1 is dedicated to $H_2O$). F1 goes through a first heater (not visible because it is at the back of the reactor) and reaches the first mixing stage M1, where it mixes with F2 and F3. The mixed stream after M1 is mixed with streams F4 and F5 in the mixing stage M2. T1, T2, and T3 show the position of the temperature measurements reported in Table 1 and 2. After M2, the stream flows through a reheater and subsequently a heat exchanger before collection.

**Figure 2:** XRD patterns of the samples produced in conditions specified in Table 1. (A) NiCo phases, (B) NiFe phases, (C) and (D) analysis of the cell volume for the brucite products.

**Figure 3:** XPS characterization of NiCo1, NiCo9, NiFe1 and NiFe2 samples. XPS spectra of Ni 2p in NiCo1 (A), Co 2p in NiCo1 (B), Ni 2p in NiCo9 (C), Co 2p in NiCo9 (D), Ni 2p in NiFe1 (E), Fe 2p in NiFe1 (F), Ni 2p in NiFe2 (G), and Fe 2p in NiFe2 (H).

**Figure 4:** X-ray photoelectron spectra of O 1s for NiCo1, NiCo9, NiFe1 and NiFe2.

**Figure 5:** TEM characterization of the NiCo1 and NiCo9 samples. STEM image of an aggregate of NiCo1 hydroxide platelets (A) and relative EDS map (B). STEM image of a single NiCo1 platelet (C) and relative EDS map (D), Ni in green, Co in blue. TEM image of NiCo9 (E) and corresponding SAED analysis.

**Figure 6:** TEM characterization of the NiFe2 sample. (A) TEM image of several particles and relative electron diffraction pattern in the inset, TEM image of a single particle and relative electron diffraction pattern (B), STEM image (C) and relative EDS map (D). Ni in green, Fe in orange.

**Figure 7**: model of the crystalline structure of NiCo1 (A), NiCo9 (B), and NiFe2 (C). In the models: red spheres O, green spheres Ni, blue spheres Co, orange spheres Fe and light blue spheres intercalated species. Scheme summarizing the result obtained (D).

**Figure 1**

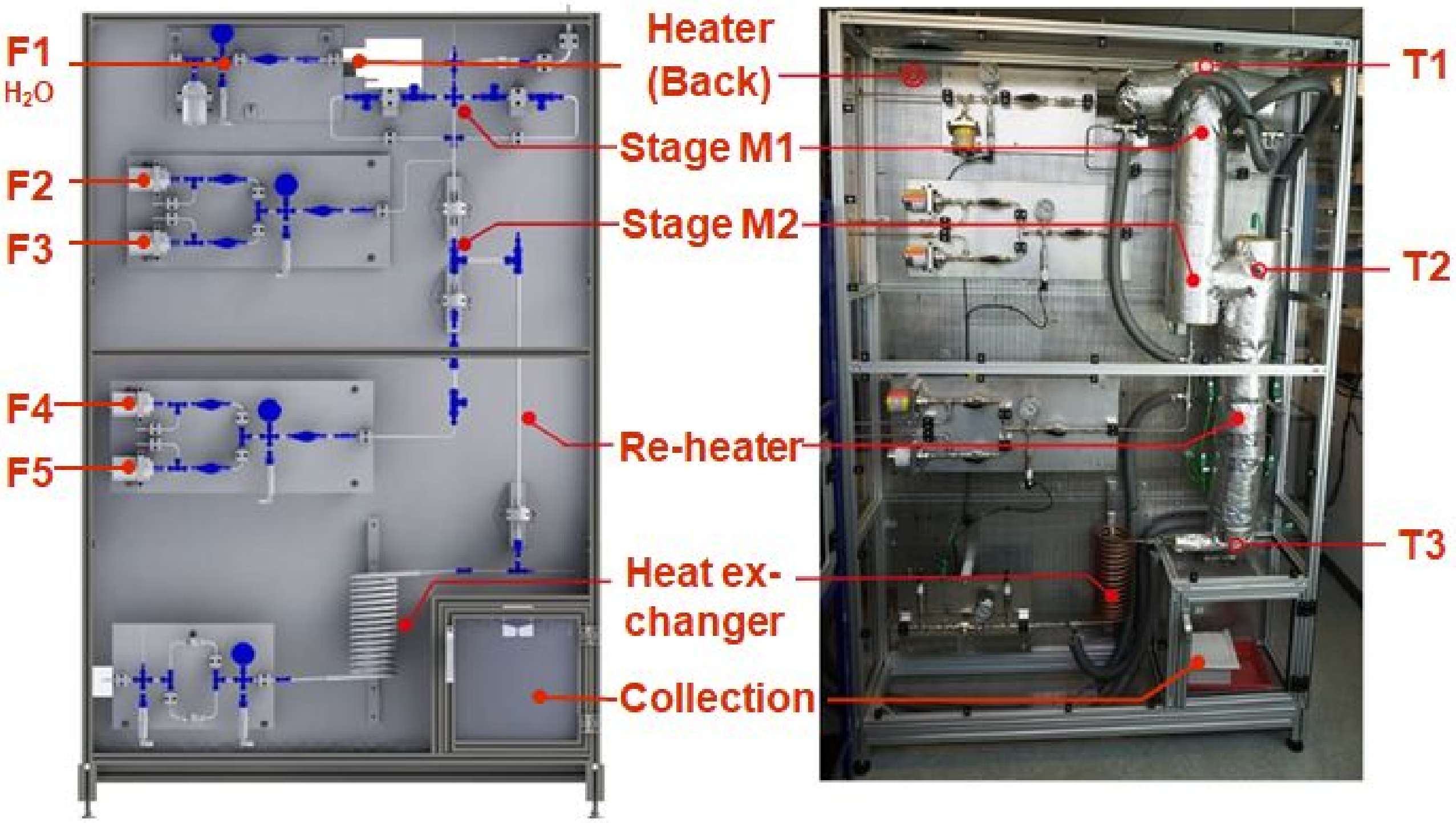

**Figure 2**

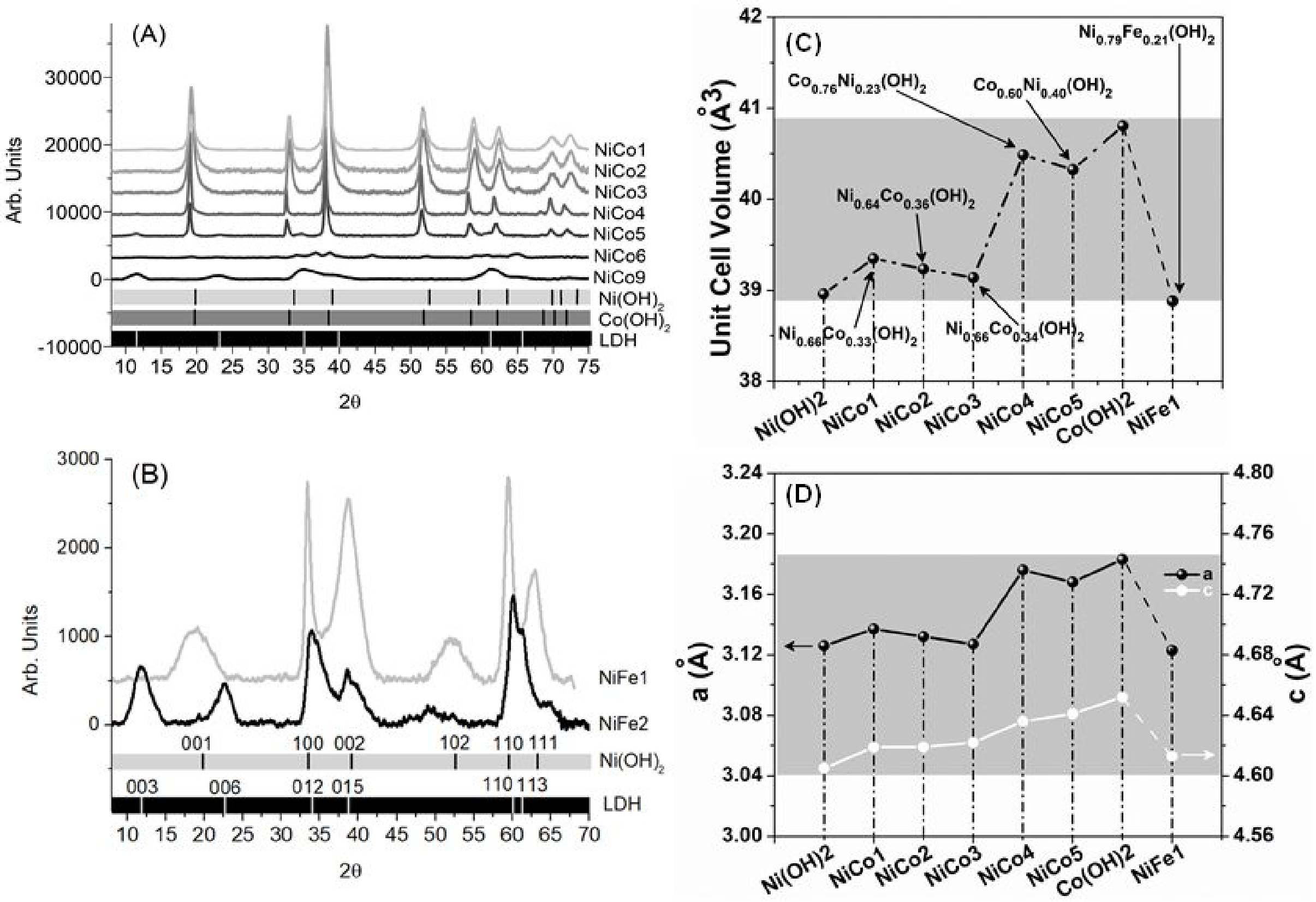

**Figure 3**

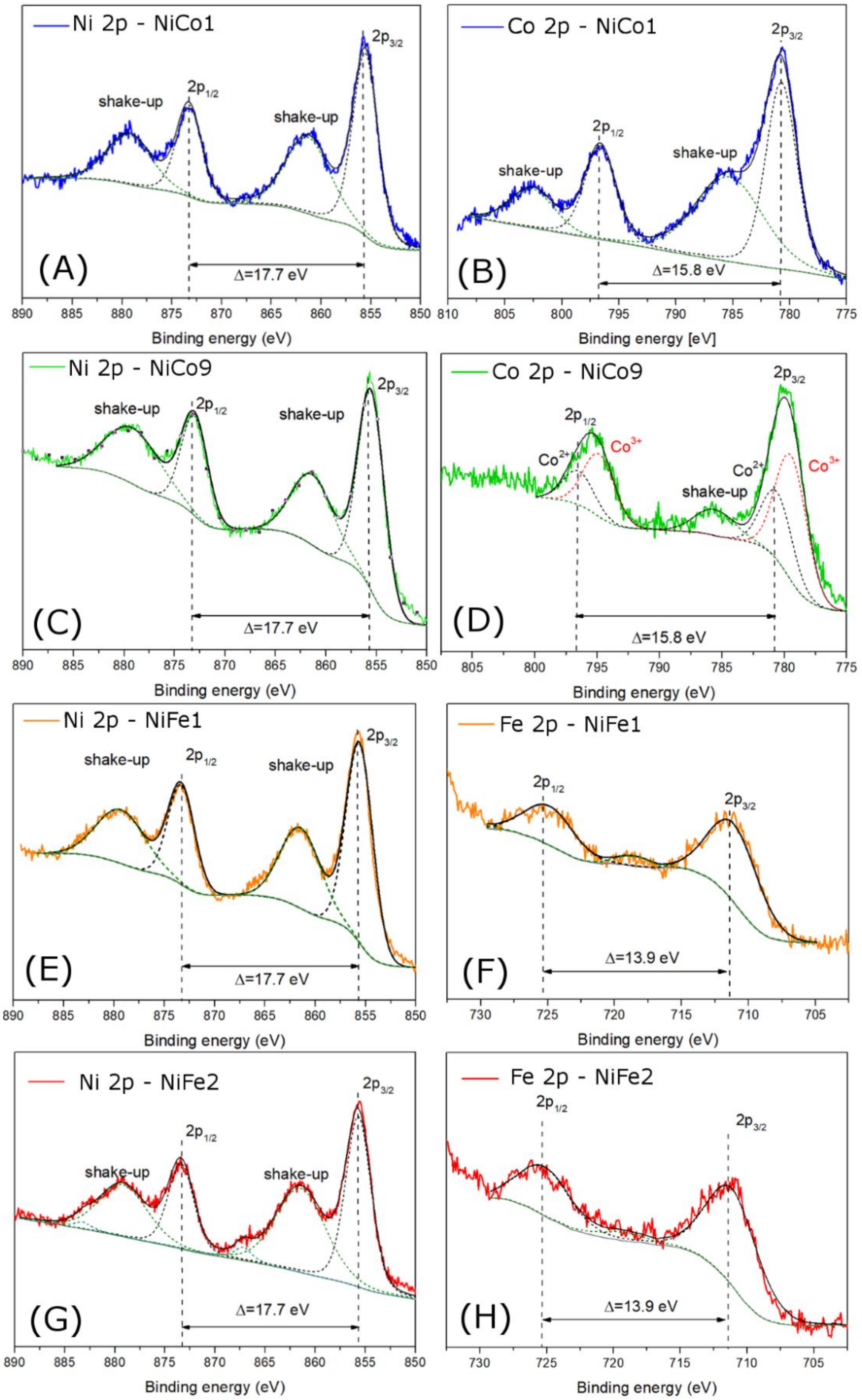

**Figure 4**

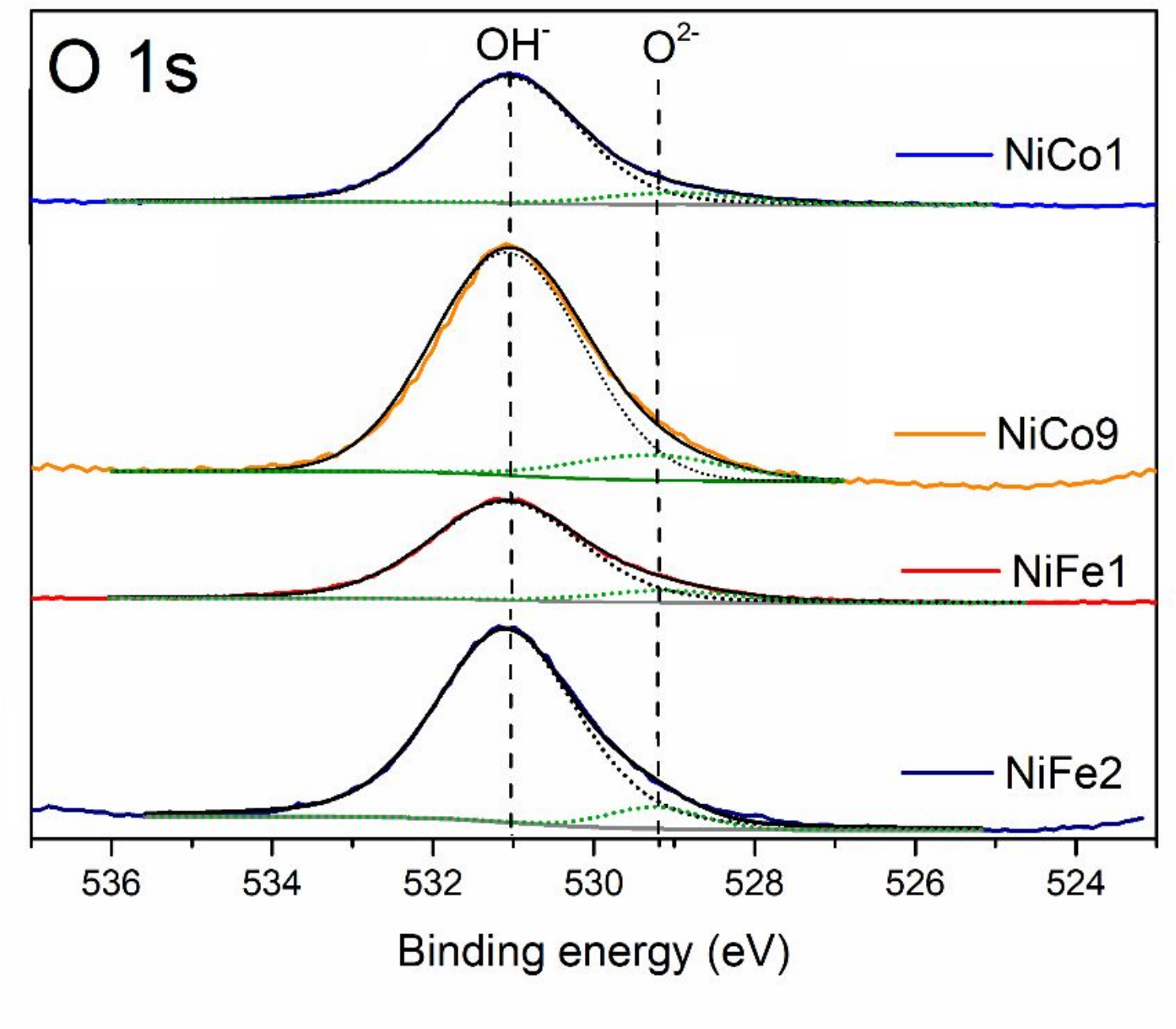

**Figure 5**

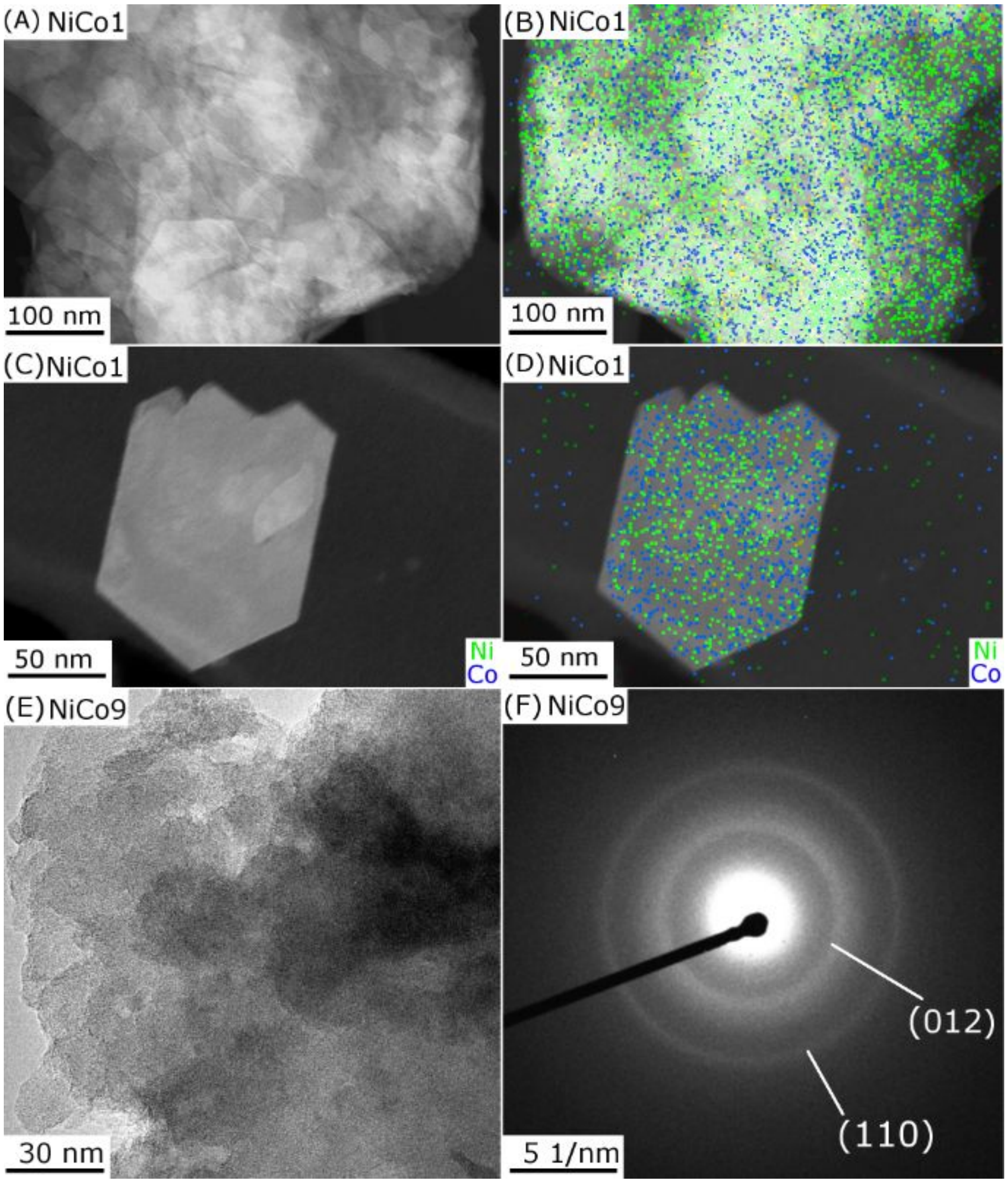

**Figure 6**

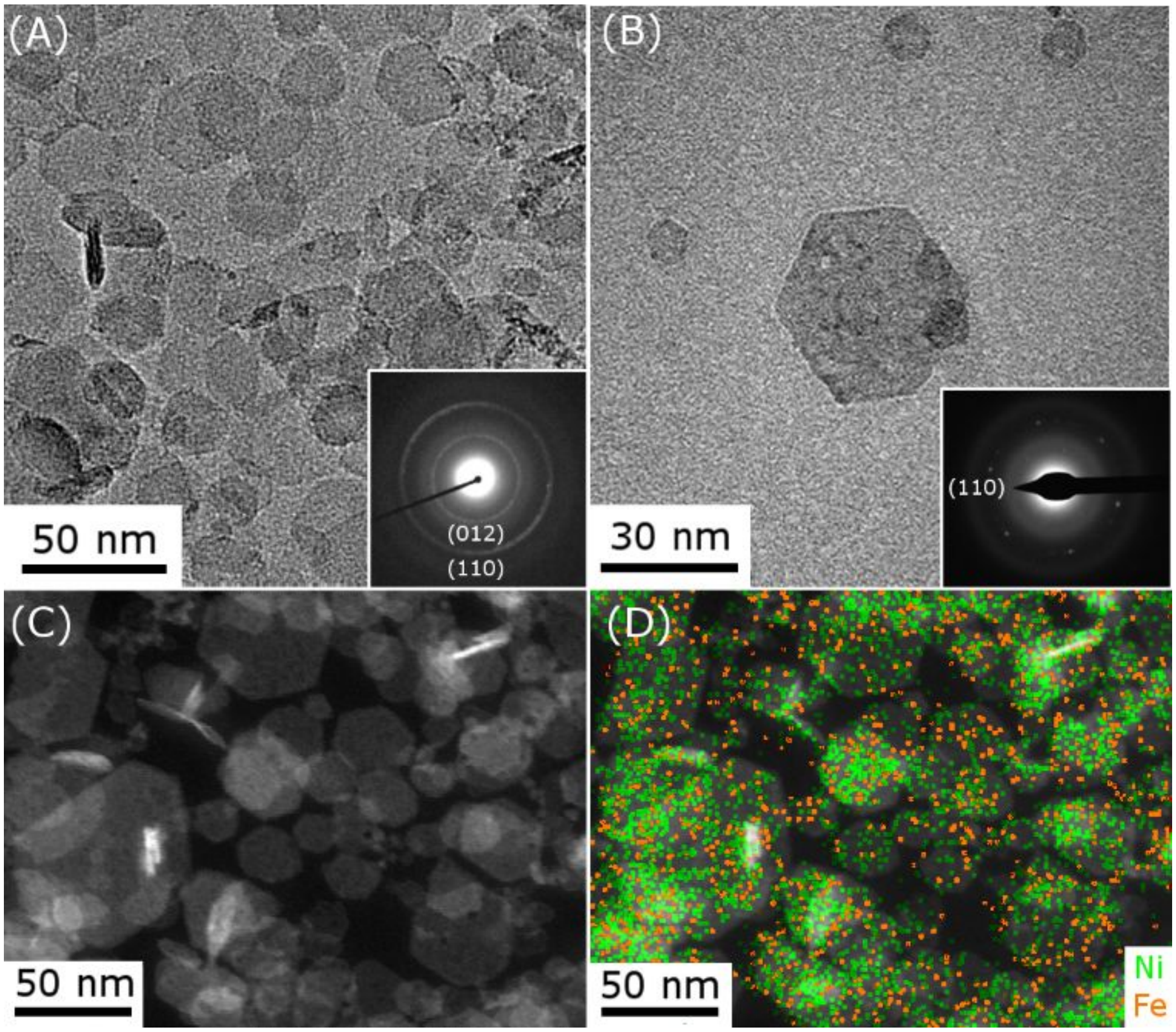

**Figure 7**

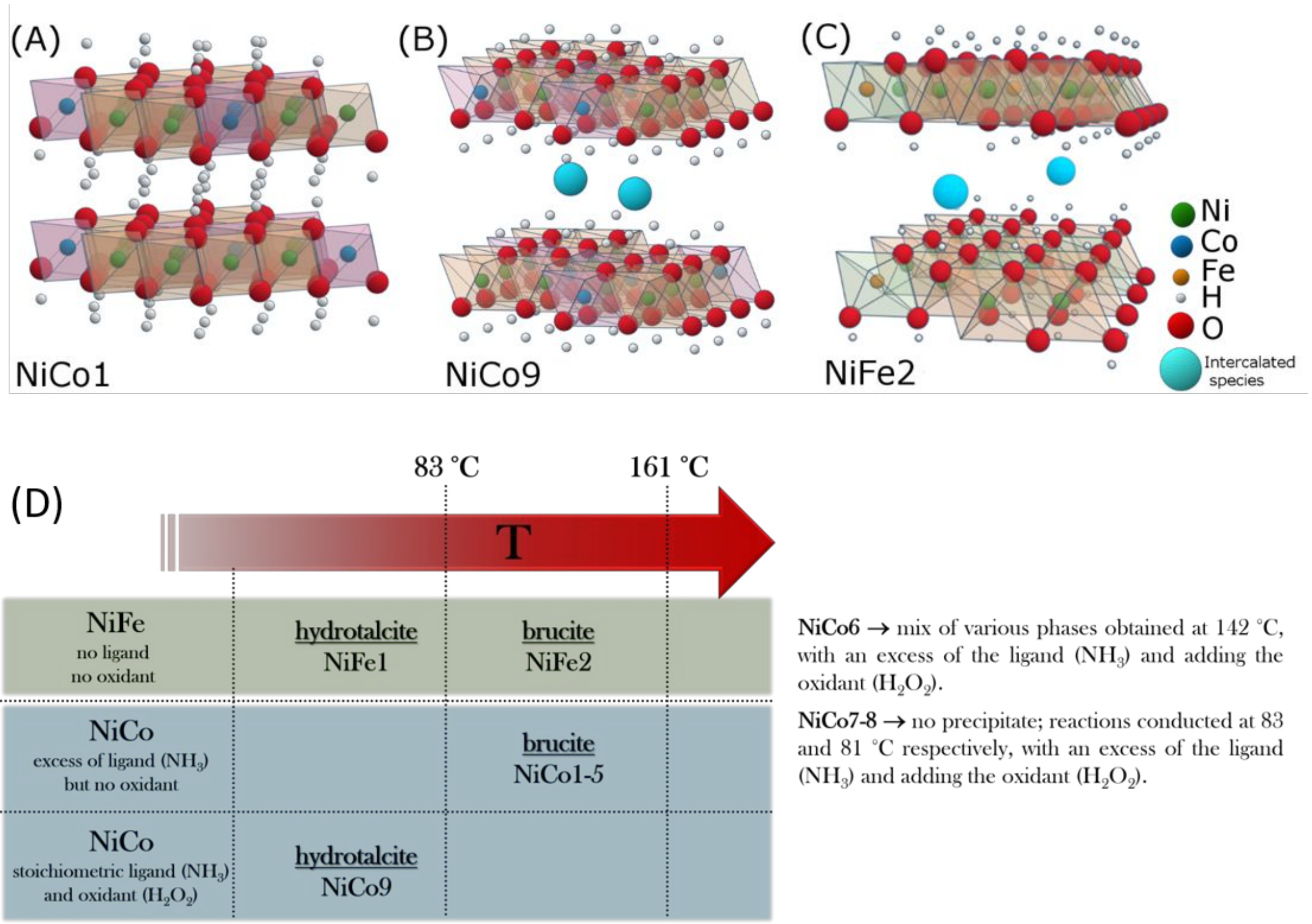

(A)
NiCo1
(B)
NiCo9
(C)
NiFe2
Ni
Co
Fe
H
O
Intercalated species
(D)
83 °C
161 °C
T
NiFe
no ligand
no oxidant
hydrotalcite
NiFe1
brucite
NiFe2
NiCo
excess of ligand (NH3)
but no oxidant
brucite
NiCo1-5
NiCo
stoichiometric ligand (NH3)
and oxidant (H2O2)
hydrotalcite
NiCo9
NiCo6 → mix of various phases obtained at 142 °C, with an excess of the ligand (NH3) and adding the oxidant (H2O2).
NiCo7-8 → no precipitate; reactions conducted at 83 and 81 °C respectively, with an excess of the ligand (NH3) and adding the oxidant (H2O2).